\documentclass{revtex4}
\usepackage{amssymb}
\usepackage{amsmath}
\usepackage[dvipsone]{graphics}
\usepackage{epsfig}
\usepackage{bm}
\usepackage{natbib}
\usepackage{color}

\fontsize{12}{12}

\expandafter\ifx\csname package@font\endcsname\relax\else
\expandafter\expandafter
 \expandafter\usepackage
 \expandafter\expandafter
 \expandafter{\csname package@font\endcsname}%
\fi

\newcommand{\be}{\begin{equation}}
\newcommand{\en}{\end{equation}}
\newcommand{\br}{\begin{eqnarray}}
\newcommand{\er}{\end{eqnarray}}

\renewcommand{\l}{\label}

\newcommand{\sss}{\scriptscriptstyle}

\renewcommand{\le}{\left}
\renewcommand{\r}{\right}
\def\ct#1{\cite{#1}}

\def\Quadrat#1#2{{\vcenter{\hrule height #2
  \hbox{\vrule width #2 height #1 \kern#1
    \vrule width #2}
  \hrule height #2}}}
\def\dAl{\mathop{\kern 1pt\hbox{$\Quadrat{8pt}{0.4pt}$} \kern1pt}}

\begin{document}
\title{Millimeter Laser Ranging to the Moon: a comprehensive theoretical model for advanced data analysis }
\author{\bf Sergei M. Kopeikin}
\affiliation{Department of Physics and Astronomy, University of Missouri-Columbia,  MO 65211, USA\\
kopeikins@missouri.edu/Fax:+1-573-882-4195}
\maketitle
\noindent
{\bf Abstract.} {\it Lunar Laser Ranging (LLR) measurements are crucial for advanced exploration of the evolutionary history of the lunar orbit, the laws of fundamental gravitational physics, selenophysics and geophysics as well as for future human missions to the Moon. Current LLR technique measures distance to the Moon with a precision approaching 1 millimeter that strongly demands further significant improvement of the theoretical model of the orbital and rotational dynamics of the Earth-Moon system. This model should inevitably be based on the theory of general relativity, fully incorporate the relevant geophysical/selenophysical processes and rely upon the most recent IAU standards.
We discuss new methods and approaches in developing such a mathematical model. The model takes into account all classic and relativistic effects in the orbital and rotational motion of the Moon and Earth at the millimeter-range level. It utilizes the IAU 2000 resolutions on reference frames and demonstrates how to eliminate from the data analysis all spurious, coordinate-dependent relativistic effects playing no role in selenophysics/geophysics. The new model is based on both the locally-inertial and barycentric coordinates and extends the currently used LLR code. The new theory and the millimeter LLR will give us the opportunity to perform the most precise fundamental test of general relativity in the solar system in robust and physically-adequate way.}

\section*{Lunar Laser Ranging}\l{back}
The dynamical modeling for the solar system (major and minor planets), for deep space navigation, and for the dynamics of Earth's satellites and the Moon must be consistent with general relativity. Lunar laser ranging (LLR) measurements are particularly crucial for testing general relativistic predictions and advanced exploration of other laws of fundamental gravitational physics. Current LLR technologies allow us to arrange the measurement of the distance from a laser on the Earth to a corner-cube reflector (CCR) on the Moon with a precision approaching 1 millimeter \cite{battat,murthyetal08}. There is a proposal to place a new CCR array on the Moon \cite{llr21century}, and possibly to install other devices such as microwave transponders \cite{transponder} for multiple scientific and technical purposes.
Successful human exploration of the Moon strongly demands further significant improvement of the theoretical model of the orbital and rotational dynamics of the Earth-Moon system. This model should inevitably be based on the theory of general relativity, fully incorporate the relevant geophysical processes, lunar libration, tides, and should rely upon the most recent standards and recommendations of the IAU for data analysis \cite{iau2000}.

LLR technique is currently the most
effective way to study the interior of the Moon and dynamics
of the Moon-Earth system. The most important contributions
from LLR include: detection of a molten lunar core 
and measurement of its influence on the Moon's orientation along with tidal dissipation \ct{moondissipation,mooncore,moonfluidcore,williams2007}; detection of lunar free libration
along with the forced terms from Venus \ct{moonlibration} and the internal excitation mechanisms \ct{libration}; an accurate test of the strong principle of equivalence for
massive bodies  \ct{sep1976,sep1998} also known
as the Nordtvedt effect \cite[Section 8.1]{willbook};
and setting of a stringent limit on time variability of the universal
gravitational constant and (non)existence of long-range
fields besides the metric tensor \ct{2001LNP...562..317N}.
LLR analysis has also given access to more subtle tests of
relativity \ct{1991ApJ...382L.101M,1996PhRvD..54.5927M,test2004,2008PhRvD..78b4033S},
measurements of the Moon's tidal acceleration \ct{1978Sci...166..977C,1994AnShO..15..129X,2002A&A...387..700C} and geodetic
precession of the lunar orbit \ct{1987PhRvL..58.1062B,1989AdSpR...9S..75D}, and has provided orders-of-magnitude improvements
in the accuracy of the lunar ephemeris \ct{1996IAUS..172...37N,2007HiA....14..472K,pmoe,2008AdSpR..42.1378K,2008AIPC..977..254S} and its three-dimensional
rotation \ct{1999A&A...343..624C,2003LPI....34.1161W}. On the geodesy front, LLR contributes
to the determination of Earth orientation parameters,
such as nutation, precession (including relativistic geodetic
precession), polar motion, UT1, and to the
long-term variation of these effects \ct{2008ASSL..349..457M,2008JGeod..82..133M}. 

LLR also contributes to
the realization of both the terrestrial and selenodesic reference
frames \ct{1996AnShO......169H,1999JGeod..73..125H}. The satellite laser ranging (SLR) realization of a dynamically-defined
inertial reference frame \ct{1990IAUS..141..173S} in contrast to the kinematically-realized frame of VLBI \cite[Section 6]{2000afce.conf.....W}, offers new possibilities for mutual
cross-checking and confirmation \ct{2008JGeod..82..133M} especially after the International Laser Ranging Service (ILRS) was established in September 1998 to support programs in
geodetic, geophysical, and lunar research activities and to provide the International Earth Rotation Service (IERS)
with products important to the maintenance of an accurate International Terrestrial Reference Frame (ITRF) \cite{ilrs02}.

Over the years, LLR has benefited from a number of
improvements both in observing technology and data modeling \ct{2002SPIE.4546..154M}. Recently,
sub-centimeter precision in determining range distances
between a laser on Earth and a retro-reflector on the Moon
has been achieved \ct{battat,murthyetal08}. As precision of LLR
measurements was gradually improving over years, enormous progress in understanding
evolutionary history of the Earth–-Moon orbit and
the internal structure of both planets has been achieved.
With the precision approaching 1 millimeter and better, accumulation
of more accurate LLR data will lead to new, fascinating
discoveries in fundamental gravitational theory, geophysics, and physics of lunar interior \ct{2007IJMPD..16.2127M} whose unique interpretation will intimately rely upon our ability to develop a systematic theoretical approach to analyze the sub-centimeter LLR data \ct{2008AdSpR..42.1378K,2009arXiv0902.2416K}.

\section*{EIH Equations of Motion in N-body Problem}\l{eqmo}

Nowadays, the theory of the lunar motion
should incorporate not only the numerous Newtonian perturbations but has to deal with much
more subtle relativistic phenomena being currently incorporated to the ephemeris codes \ct{chatc,sta,epm,pmoe}. Theoretical approach, used for construction of the ephemerides, accepts that the post-Newtonian description of the planetary motions can be achieved with the Einstein-Infeld-Hoffmann (EIH) equations of motion of point-like masses \ct{eih}, which are valid in the barycentric frame of the solar system with time coordinate, $t$, and spatial coordinates, $x^i\equiv{\bm x}$.

Due to the covariant nature of general theory of relativity the barycentric coordinates are not unique and are defined up to the space-time transformation \ct{vab,brum,sof89}
\br\l{gt1}
t&\mapsto& t-\frac{1}{c^2}\sum_{\sss{B}}\nu_{\sss{B}}\frac{G{\rm M}_{\sss{B}}}{R_{\sss{B}}}\le({\bm R}_{\sss{B}}\cdot{\bm v}_{\sss{B}}\r)\;,
\\\l{gt2}
{\bm x}&\mapsto&{\bm x}-\frac{1}{c^2}\sum_{\sss{B}}\lambda_{\sss{B}}\frac{G{\rm M}_{\sss{B}}}{R_{\sss{B}}}{\bm R}_{\sss{B}}\;,
\er
where summation goes over all the massive bodies of the solar system ($B=1,2,...,N$); $G$ is the universal gravitational constant; $c$ is the fundamental speed in the Minkowskian space-time; a dot between any spatial vectors, ${\bm a}\cdot{\bm b}$, denotes an Euclidean dot product of two vectors ${\bm a}$ and ${\bm b}$;  ${\rm M}_{\sss {B}}$ is mass of a body B; ${\bm x}_{\sss{B}}={\bm x}_{\sss{B}}(t)$ and ${\bm v}_{\sss{B}}={\bm v}_{\sss{B}}(t)$ are coordinates and velocity of the center of mass of the body B; ${\bm R}_{\sss{B}}={\bm x}-{\bm x}_{\sss{B}}$ is a relative distance from  a field point ${\bm x}$ to the body B;
$\nu_{\sss B}$ and $\lambda_{\sss{B}}$ are constant, but otherwise free parameters being responsible for a particular choice of the barycentric coordinates. We emphasize that these parameters can be chosen arbitrary for each body B of the solar system. Physically, it means that the space-time around each body is covered locally by its own coordinate grid, which matches smoothly with the other coordinate charts of the massive bodies in the buffer domain, where the different coordinates overlap.

If the bodies in N-body problem are numbered by indices $B, C, D,$ etc., and the coordinate freedom is described by equations (\ref{gt1})--(\ref{gt2}), EIH equations of motion for the body $B$ have the following form \ct{brum}
\br\label{eh4}
a^i_{\sss{B}}&=&\sum_{\sss{C\neq B}}\left[E^i_{\sss{BC}}+\frac{4-2\lambda_{\sss C}}{c}\left({\bm v}_{\sss B}\times{\bm H}_{\sss{BC}}\right)^i-\frac{3-2\lambda_{\sss C}}{c}\left({\bm v}_{\sss C}\times{\bm H}_{\sss{BC}}\right)^i\right]
\er
where $E^i_{\sss{BC}}$ is called the gravitoelectric force, and the terms associated with the cross products $\left({\bm v}_{\sss B}\times{\bm H}_{\sss{BC}}\right)^i$  and $\left({\bm v}_{\sss C}\times{\bm H}_{\sss{BC}}\right)^i$ are referred to as the gravitomagnetic force \ct{1988IJTP...27.1395N}.

The gravitoelectric force is given by
\br\l{eh5}
E^i_{\sss{BC}}&=&\left(-\frac{G{\rm M}_{\sss C}}{R^3_{\sss{BC}}}+\frac{G{\rm M}_{\sss C}}{c^2}{\cal E}_{\sss{BC}}\right)R^i_{\sss{BC}}\;,
\er
and the gravitomagnetic force is 
\br\l{eh7}
H^i_{\sss BC}&=&-\frac1{c}\left({\bm V}_{\sss{BC}}\times{\bm E}_{\sss{BC}}\right)^i=
\frac{G{\rm M}_{\sss C}}{c}\frac{\left({\bm V}_{\sss{BC}}\times{\bm R}_{\sss{BC}}\right)^i}{R^3_{\sss{BC}}}\;,
\er
where ${\bm V}_{\sss{BC}}={\bm v}_{\sss{B}}-{\bm v}_{\sss{C}}$. 
The first term in right side of equation (\ref{eh5}) is the Newtonian force of gravity, and the post-Newtonian correction
\br\l{eh6}
{\cal E}_{\sss BC}&=&-
\frac{1}{R^3_{\sss{BC}}}\Biggl\{3(-1+\lambda_{\sss{C}}) v^2_{\sss{B}}+3(1-2\lambda_{\sss{C}})(\bm v_{\sss{B}}
\cdot\bm v_{\sss C})-(1-3\lambda_{\sss{C}})v^2_{\sss C}
\\\nonumber&&-\frac32\left(\frac{\bm R_{\sss{BC}}\cdot\bm v_{\sss C}}{R_{\sss{BC}}}\right)^2
-3\lambda_{\sss{C}}\left(\frac{\bm R_{\sss{BC}}\cdot{\bm V}_{\sss{BC}}}{R_{\sss{BC}}}\right)^2
-(5-2\lambda_{\sss{B}})\frac{G{\rm M}_{\sss{B}}}{R_{\sss{BC}}}
-(4-2\lambda_{\sss{C}})\frac{G{\rm M}_{\sss C}}{R_{\sss{BC}}}
\\\nonumber&&
+\sum_{\sss{D\neq B,C}}G{\rm M}_{\sss D}\le(\frac{1+2\lambda_{\sss{C}}}{2R^3_{\sss{CD}}}-\frac{\lambda_{\sss{C}}}{R^3_{\sss{BD}}}+
\frac{3\lambda_{\sss{D}}}{R_{\sss{BD}}R^2_{\sss{BC}}}-\frac{3\lambda_{\sss{D}}}{R_{\sss{CD}}R^2_{\sss{BC}}}\r)(\bm R_{\sss{BC}}\cdot\bm R_{\sss{BD}})\Biggr\}
\\\nonumber&&
+\sum_{\sss{D\neq B,C}}G{\rm M}_{\sss D}\left[\frac{4}{R_{\sss{BC}}R^3_{\sss{CD}}}+\frac{\lambda_{\sss C}}{R_{\sss{BC}}R^3_{\sss{BD}}}-\frac{\lambda_{\sss C}}{R_{\sss{CD}}R^3_{\sss{BD}}}-\frac{7-2\lambda_{\sss D}}{2R_{\sss{BD}}R^3_{\sss{CD}}}+\frac{1-2\lambda_{\sss D}}{R^3_{\sss{BC}}R_{\sss{CD}}}
+\frac{4-\lambda_{\sss{D}}}{R^3_{\sss{BC}}R_{\sss{BD}}}    \right]\;,
\er
As one can see, the gravitomagnetic force (\ref{eh7}) is proportional to the gravitoelectric force (\ref{eh5}) multiplied by the factor of $V/c$, where $V$ is the relative velocity between two gravitating bodies.
EIH equations (\ref{eh4})--(\ref{eh7}) differ from the equations of the PPN formalism {1969ApJ...158...81E} employed in particular at JPL for actual calculation of the ephemerides of the major planets by the fact that the right side of equation (\ref{eh4}) has been resolved into radius-vectors and velocities of the massive bodies and does not contain second derivatives (accelerations). This elimination of the high-order time derivatives from a perturbed force is a standard practice in celestial mechanics for calculation of the perturbed motion.

Barycentric coordinates ${\bm x}_{\sss{B}}$ and velocities ${\bm v}_{\sss{B}}$ of the center of mass of body $B$ are adequate theoretical quantities for description of the world-line of the body with respect to the center of mass of the solar system. However, the barycentric coordinates are global coordinates covering the entire solar system. Therefore, they have little help for efficient physical decoupling of the post-Newtonian effects existing in the orbital and rotational motions of a planet and for the description of motion of planetary satellites around the planet. The problem stems from the covariant nature of EIH equations, which originates from the fundamental structure of space-time manifold and the gauge freedom of the general relativity theory.

This freedom is already seen in the post-Newtonian EIH force (\ref{eh4})--(\ref{eh6}) as it explicitly depends on the choice of spatial coordinates through parameters $\lambda_{\sss C}, \lambda_{\sss D}$. Each term depending explicitly on $\lambda_{\sss C}, \lambda_{\sss D}$ in equations (\ref{eh4})--(\ref{eh6}), has no direct physical meaning as it can be eliminated after making a specific choice of these parameters. In many works on experimental gravity and applied relativity researches fix parameters $\lambda_{\sss C}=\lambda_{\sss D}=0$, which corresponds to working in harmonic coordinates. Harmonic coordinates simplify EIH equations to large extent but one has to keep in mind that they have no physical privilege anyway, and that a separate term or a limited number of terms from EIH equations of motion can not be measured -- only coordinate-independent effects can be measured \ct{brum}.

Recently, there was a lot of discussions about whether LLR can measure the gravitomagnetic field $H^i_{\sss BC}$ \ct{2007PhRvL..98g1102M,2007PhRvL..98v9002M,test2004,k07,2008PhRvD..78b4033S}. The answer to this question is subtle and requires more profound theoretical consideration involving the process of propagation of the laser pulses in a curved space-time of the Earth-Moon system. We are hoping to discuss this topic somewhere else.  Nevertheless, what is evident already now is that equation (\ref{eh4}) demonstrates a strong dependence of the gravitomagnetic force on the choice of coordinates. For this reason, by changing the coordinate parameter $\lambda_{\sss C}$ one can eliminate either the term $\left({\bm v}_{\sss B}\times{\bm H}_{\sss{BC}}\right)^i$  or $\left({\bm v}_{\sss C}\times{\bm H}_{\sss{BC}}\right)^i$ from EIH equations of motion (\ref{eh4}). It shows that the strength of the factual gravitomagnetic force, as it appears in the equations of motion, is coordinate-dependent, and, hence, a great care should be taken in order to properly interpret the LLR "measurement" of such gravitomagnetic terms in consistency with the covariant nature of the general theory of relativity and the theory of astronomical measurements in curved space-time \ct{1981rcse.conf..283B,1962rdgr.book..441S, syngebook,inpl,vab}.

\section*{The Gauge Freedom}\l{rgfr}

The primary gauge freedom of EIH equations of motion is associated with the transformations (\ref{gt1})--(\ref{gt2}) of the barycentric coordinates of the solar system, which are parameterized by parameters $\nu_{\sss C}$ and $\lambda_{\sss C}$. However, the post-Newtonian force in the lunar equations of motion admits additional freedom of coordinate transformations. This {\it residual} freedom remains even after fixing the coordinate parameters $\nu_{\sss C}$ and $\lambda_{\sss C}$ in equations ({\ref{eh4})--(\ref{eh6}).  It is associated with the fact that the Earth-Moon system moves in tidal gravitational field of the Sun and other planets, which presence in the local frame of the Earth-Moon system indicates that the local background space-time is not asymptotically-flat. The residual freedom remains in making transformations of the local coordinates attached to the Earth-Moon system. It induces the gauge transformation of the metric tensor and the Christoffel symbols and changes the structure of the post-Newtonian terms in EIH equations of motion of the Earth-Moon system that is not associated with the parameters $\nu_{\sss C}$ and $\lambda_{\sss C}$.

Thus, we face the problem of investigation of the residual gauge freedom of the lunar equations of motion, which goes beyond the choice of the barycentric coordinates by fixing a specific value of the gauge parameter $\lambda_{\sss C}$ in equations (\ref{eh4})--(\ref{eh6}). This freedom is naturally associated with the choice of the local coordinates of the Earth-Moon barycentric frame as well as the geocentric and selenocentric reference frames. Proper choice of the local coordinates removes all non-physical degrees of freedom from the metric tensor and eliminates spurious (non-measurable) terms from the post-Newtonian forces in the equations of relative motion of the Moon with respect to the Earth. If one ignores the residual gauge freedom, the gauge-dependent terms will infiltrate the equations of motion causing possible misinterpretation of LLR observations. This problem is similar to that one meets in
cosmology, where the theory of cosmological perturbations is designed essentially in terms of the gauge-independent variables so that observations of various cosmological effects are not corrupted by the spurious, coordinate-dependent signals \ct{1992PhR...215..203M}. The residual gauge degrees of freedom existing in the relativistic three-body problem (Sun-Earth-Moon), can lead to misinterpretation of various aspects of gravitational physics of the Earth-Moon system \ct{2008arXiv0809.3392K,k07}, thus, degrading the value of extremely accurate LLR measurements for testing fundamental physics of space-time and deeper exploration of the lunar interior \ct{2008AdSpR..42.1378K}.

The residual gauge freedom of the three body problem  was studied by Brumberg and Kopeikin \ct{bk-nc}, Klioner and Voinov \ct{klv}, and Damour, Soffel and Xu \ct{dsx4}. They found that the post-Newtonian equations of motion of a test body (artificial satellite) can be significantly simplified by making use of a four-dimensional space-time transformation from the solar barycentric coordinates $x^\alpha=(ct, {\bm x})$, to the geocentric coordinates $X^\alpha=(cT, {\bm X})$
\br\l{gct1}
T&=&t+\frac1{c^2}A(t,{\bm r}_{\sss E})+\frac1{c^4}B(t,{\bm r}_{\sss E})+O\le(\frac1{c^5}\r)\;,
\\\l{gct2}
X^i&=&x^i-x^i_{\sss E}(t)+\frac1{c^2}C^i(t,{\bm r}_{\sss E})+O\le(\frac1{c^4}\r)\;,
\er
where the gauge functions $A(t,{\bm x})$, $B(t,{\bm x})$, $C^i(t,{\bm x})$ are polynomials of the geocentric distance ${\bm r}_{\sss E}={\bm x}-{\bm x}_{\sss E}(t)$ of the field point ${\bm x}$ from the Earth's geocenter, ${\bm x}_{\sss E}(t)$. Coefficients of these polynomials are functions of the barycentric time $t$ that are determined by solving a system of ordinary differential equations, which follow from the gravity field equations and the tensor law of transformation of the metric tensor from one coordinate chart to another \ct{1988CeMec..44...87K}.
Contrary to the test particle, the Moon is a massive body, which makes the exploration of the residual gauge freedom of the lunar motion more involved. This requires introduction of one global (SSB) frame and three local reference frames associated with the Earth-Moon barycenter, the geocenter, and the center of mass of the Moon (selenocenter). It should be clearly understood that any coordinate system can be used for processing and interpretation of LLR data since any viable theory of gravity obeys the Einstein principle of relativity, according to which there is no preferred frame of reference \ct{lali,fockbook,mtw}. Accepting the Einstein principle of relativity leads to discarding any theory of gravity based on a privileged frame (aether) \ct{2006dclw.conf..163E} or admitting a violation of the Lorentz invariance \ct{2008APS..DMP.I3001K}. The class of scalar-tensor theories of gravity, which have two PPN parameters - $\beta$ and $\gamma$ \cite{willbook,1992CQGra...9.2093D}, is in agreement with the principle of relativity and it has been investigated fairly well \ct{kovl,2000PhRvD..62b4019K}.

The principle of relativity also assumes that an arbitrary chosen, separate term in the post-Newtonian equations of motion of massive bodies can not be physically interpreted as straightforward as in the Newtonian physics. The reason is that the post-Newtonian transformations (\ref{gt1})--(\ref{gt2}) and (\ref{gct1})--(\ref{gct2}) of the barycentric and local coordinates, change the form of the equations of motion so that they are not form-invariant. Therefore, only those post-Newtonian effects, which do not depend on the frame transformations can have direct physical interpretation. For example, the gauge parameters $\nu_{\sss C}$ and $\lambda_{\sss C}$ entering transformations (\ref{gt1})--(\ref{gt2}) and EIH equations (\ref{eh4})--(\ref{eh7}) can not be determined from LLR data irrespectively of their accuracy because these parameters define the barycentric coordinates and can be fixed arbitrary by observer without any relation to observations.

\section*{Towards a New Lunar Ephemeris}\l{reh}

Existing computer-based theories of the lunar ephemeris \ct{chatc,sta,epm,pmoe} consist of three major blocks:
\begin{enumerate}
\item[(1)] the barycentric EIH equations (\ref{eh4})--(\ref{eh6}) of orbital motion of the Moon, Earth, Sun, and other planets of the solar system with the gauge parameters $\nu_{\sss C}=1/2$,  $\lambda_{\sss C}=0$ - the standard PPN coordinates;
\item[(2)] the Newtonian rotational equations of motion of the Moon and Earth;
\item[(3)] the barycentric post-Newtonian equations of motion for light rays propagating from laser to CCR on the Moon and back in standard coordinates with the gauge parameters $\nu_{\sss C}=1/2$,   $\lambda_{\sss C}=0$.
\end{enumerate}
This approach is straightforward but it does not control gauge-dependent terms in EIH equations of motion associated with the choice of the gauge-fixing parameters $\nu_{\sss C}$ and $\lambda_{\sss C}$. Particular disadvantage of the barycentric approach in application to the lunar ephemeris is that it mixes up the post-Newtonian effects associated with the orbital motion of the Earth-Moon barycenter around the Sun with those, which are attributed exclusively to the relative motion of the Moon around the Earth. This difficulty is also accredited to the gauge freedom of the equations of motion in three-body problem and was pointed out in papers \ct{bk-nc,dsx4,2000A&A...363..335T}. Unambiguous decoupling of the orbital motion of the Earth-Moon barycenter from the relative motion of the Moon around the Earth with apparent identification of the gauge-dependent degrees of freedom in the metric tensor and equations of motion is highly desirable in order to make the theory more sensible and to clean up the LLR data processing software from the fictitious coordinate-dependent perturbations, which do not carry out any physically-relevant information and may accumulate errors in numerical ephemerides of the Moon and the Earth.

This goal can be rationally achieved if the post-Newtonian theory of the lunar motion is consistently extended to account for mathematical properties offered by the scalar-tensor theory of gravity and the differential structure of the space-time manifold. Altogether it leads us to the idea that besides the global barycentric coordinates of the solar system one has to introduce three other local reference frames. The origin of these frames should be fixed at the Earth-Moon system barycenter, the Earth's center of mass (geocenter), and the Moon's center of mass (selenocenter). We distinguish the Earth-Moon barycenter from the geocenter because the Moon is not a test particle, thus, making the Earth-Moon barycenter displaced from the geocenter along the line connecting the Earth and Moon and located approximately 1710 km below the surface of the Earth. Mathematical construction of each frame is reduced to finding a metric tensor by means of solution of the gravity field equations with an appropriate boundary condition \ct{2009arXiv0902.2416K}. The gauge freedom of the three-body problem is explored by means of matching the set of the metric tensors defined in each reference frame in the overlapping domains of their applicability associated with the specific choice of boundary conditions imposed in each frame on the metric tensor. This matching procedure is an integral part of the equations defining the local differential structure of the manifold \ct{eisen,dfn}, which proceeds from a requirement that the overlapping space-time domains covered by the local reference frames, are diffeomorphic.

The primary objective of the multi-frame post-Newtonian theory of the lunar ephemeris is the development of a new set of analytic equations to revamp the LLR data processing software in order to suppress the spurious gauge-dependent solutions, which may overwhelm the existing barycentric code at the millimeter accuracy of LLR measurements, thus, plunging errors in the interpretation of selenophysics, geophysics and fundamental gravitational physics. Careful mathematical construction of the local frames with the post-Newtonian accuracy will allow us to pin down and correctly interpret all physical effects having classical (lunar interior, Earth geophysics, tides, asteroids, etc.) and relativistic nature. The gauge freedom in the three-body problem (Earth-Moon-Sun) has been carefully examined in our paper \ct{2009arXiv0902.2416K} by making use of a scalar-tensor theory of gravity and the principles of the analytic theory of relativistic reference frames in the solar system \ct{1988CeMec..44...87K,bk89,dsx1} that was adopted by the XXIV-th General
Assembly of the International Astronomical Union \ct{iau2000,2007AIPC..886..268K} as a standard for data processing of high-precision astronomical observations.

The advanced post-Newtonian dynamics of the Sun-Earth-Moon system must include the following structural elements:
\begin{enumerate}
\item construction of a set of astronomical reference frames decoupling orbital dynamics of the Earth-Moon system from the rotational motion of the Earth and Moon with the full account of the post-Newtonian corrections and elimination of the gauge modes;
\item relativistic definition of the integral parameters like mass, the center of mass, the multipole moments of the gravitating bodies;
\item derivation of the relativistic equations of motion of the center-of-mass of the Earth-Moon system with respect to the barycentric reference frame of the solar system;
\item derivation of the relativistic equations of motion of the Earth and Moon with respect to the reference frame of the Earth-Moon system;
\item derivation of the relativistic equations of motion of CCR on the Moon (or a lunar orbiter that is deployed with CCR) with respect to the selenocentric reference frame;
\item derivation of the relativistic equations of motion of a laser with respect to the geocentric reference frame.
\end{enumerate}
These equations must be incorporated to LLR data processing software operating with observable quantities, which are proper times of the round trip of the laser pulses between the laser on the Earth and CCR on the Moon. The computational advantage of the new approach to the lunar ephemeris is that it separates clearly physical effects from the choice of coordinates. This allows us to get robust measurement of true physical parameters of the LLR model and give them direct physical interpretation. The new approach is particularly useful for comparing different models of the lunar interior and for making the fundamental test of general theory of relativity.
\begin{acknowledgments}
This work has been supported by the Research Council Grant No. C1669103 of the University of Missouri-Columbia and by 2008-09 faculty incentive grant of the Arts and Science Alumni Organization of the University of Missouri-Columbia.
\end{acknowledgments}
\bibliography{Lunar_Motion_update3}
\end{document}